%% file: sample-sigconf.tex
\documentclass[sigconf,9pt]{acmart}

\usepackage{graphicx}
\usepackage{textcomp}
\usepackage{xcolor}
\usepackage{amsmath} 
\usepackage{url}
\usepackage[capitalise]{cleveref}
\usepackage{amsfonts}
\usepackage{color}
\usepackage[linesnumbered,ruled]{algorithm2e}
\SetKw{Continue}{continue}
\SetKw{Break}{break}
\SetKwInput{KwInput}{Input}                
\SetKwInput{KwOutput}{Output}              
\SetKwComment{Comment}{$\triangleright$\ }{}

\usepackage{longtable, array, booktabs}
\usepackage{multirow}
\AtBeginDocument{%
  \providecommand\BibTeX{{%
    \normalfont B\kern-0.5em{\scshape i\kern-0.25em b}\kern-0.8em\TeX}}}

\copyrightyear{2026}
\acmYear{2026}
\setcopyright{cc}
\setcctype{by}
\acmConference[MLCAD '26]{2026 ACM/IEEE International Symposium on Machine Learning for CAD}{September 07--09, 2026}{Jeju Island, Republic of Korea}
\acmBooktitle{2026 ACM/IEEE International Symposium on Machine Learning for CAD (MLCAD '26), September 07--09, 2026, Jeju Island, Republic of Korea}
\acmDOI{10.1145/3831599.3840311}
\acmISBN{979-8-4007-2878-5/2026/09}

\begin{document}
\begin{sloppypar}


\title{Coverage-Driven RTL Assertion Generation with Formal Exploration and Neuro-Symbolic Refinement}

\author{Zhiyuan Yan\textsuperscript{1}, Ziyue Zheng\textsuperscript{1}, and Hongce Zhang\textsuperscript{1,2}}
\affiliation{%
  \institution{\{zyan760, zzheng989\}@connect.hkust-gz.edu.cn; hongcezh@hkust-gz.edu.cn\\
  \textsuperscript{1}The Hong Kong University of Science and Technology (Guangzhou)\\
  \textsuperscript{2}The Hong Kong University of Science and Technology}
  \city{}
  \country{}
}

\renewcommand{\authors}{Zhiyuan Yan, Ziyue Zheng,  and Hongce Zhang}
\renewcommand{\shortauthors}{Yan, Zheng, and Zhang}

\input{sections/abstract}

\maketitle

\input{sections/intro}
\input{sections/background}
\input{sections/Method}
\input{sections/Experiment}
\input{sections/related-works}
\input{sections/conclusion}


\bibliographystyle{ACM-Reference-Format}
\bibliography{sample-base}
\end{sloppypar}
\end{document}

%% file: sections/abstract.tex
\begin{abstract}

Hardware functional verification relies on high-quality assertions to expose 
design bugs and establish confidence in Register Transfer Level (RTL) designs. 
Yet existing assertion mining methods still struggle to produce complete and 
reliable assertion sets: random or limited traces fail to cover hard-to-reach 
behaviors, and one-shot generation provides little feedback about what remains 
unverified or how the assertion set should be improved. As a result, critical 
design behaviors can remain uncovered even when many assertions are generated. \looseness = -1

We present NeuroAssertion, a coverage-driven assertion generation framework that combines formal 
trace generation, syntax-guided synthesis (SyGuS), and an agent-inspired refinement process within 
a unified framework. Our framework first converts hard-to-reach control-flow conditions into 
formal reachability objectives, uses model checking to generate behaviorally diverse traces, and mines 
initial assertions from these traces with SyGuS. It then performs targeted 
agent-inspired refinement under verification feedback: one LLM first proposes candidate 
assertions for uncovered regions, and if a candidate fails formal checking, a 
second LLM generates a repair grammar that guides constrained symbolic 
synthesis in a neuro-symbolic repair procedure. Experimental results show that this 
framework delivers around 2$\times$ more assertions and about 2$\times$ 
higher mutation coverage than traditional assertion mining methods. \looseness = -1
\end{abstract}

%% file: sections/intro.tex
\section{Introduction}

Hardware functional verification increasingly depends on high-quality
assertions to expose bugs and establish confidence in Register Transfer
Level (RTL) designs. A central technique is assertion-based verification
(ABV), where SystemVerilog Assertions (SVAs) capture expected design
behaviors for simulation and formal property verification. Yet writing
high-quality SVAs remains labor-intensive and error-prone, making
assertion construction a persistent bottleneck in modern verification
flows.

To reduce this manual effort, a substantial body of prior work has
studied mining assertions from simulation traces. GoldMine~\cite{vasudevan2010goldmine}
mines assertions from RTL simulation traces using static analysis and
decision-tree learning. Liu et al.~\cite{liu2011automatic,liu2012word}
extended this direction from transaction-level models to word-level
feature discovery, enabling mined assertions to capture relationships
between multi-bit signals. Sheridan et al.~\cite{sheridan2014coverage}
further emphasized coverage-aware assertion selection to reduce
redundancy. Later template-based methods, including
A-TEAM~\cite{danese2017team} and HARM~\cite{germiniani2022harm}, imposed
stronger structural priors on the mining process. Most recently,
SMART~\cite{ye2025unlocking} combined SyGuS with oracle-guided
counterexample refinement to improve trace-driven assertion synthesis.

Despite this progress, existing workflows still face two limitations.
First, mining quality is constrained by trace diversity: when initial
mining relies on random or constrained-random traces, hard-to-reach RTL
behaviors often remain unobserved, leaving the initial assertion set
incomplete. Second, existing methods provide little feedback about
assertion completeness. They mine assertions from observed traces
without explicitly identifying uncovered behaviors or how refinement
should proceed. As a result, assertion mining remains largely one-shot:
missed behaviors stay hidden, and improvement lacks direction. In
practice, however, engineers use coverage feedback to extend the
assertion set toward uncovered behaviors.

These limitations suggest that RTL assertion mining should be treated
as a coverage-driven refinement problem rather than a one-shot
generation problem. NeuroAssertion follows this view. It first uses
formal exploration to convert hard-to-reach control-flow conditions into
reachability objectives and generate traces that expose harder-to-reach
behaviors before SyGuS-based mining. It then uses mutation-coverage
feedback to identify uncovered obligations, generate LLM-based
candidate assertions, and repair invalid ones through an
agent-inspired loop with grammar-constrained symbolic synthesis.
Together, these stages expand the behavioral basis of mining and refine
the assertion set under explicit mutation-coverage feedback and formal
checking.

Overall, this paper makes the following contributions:

\begin{itemize}
    \item We propose  NeuroAssertion, a coverage-driven assertion generation framework
    that combines formal exploration, syntax-guided synthesis, and
    an agent-inspired refinement loop within a unified framework, enabling assertion
    mining to move beyond one-shot generation over limited observed
    behaviors. \looseness = -1
    
    \item We introduce an agent-inspired refinement process that uses
    mutation-coverage feedback to target uncovered obligations with
    LLM-generated candidate assertions and
    then repairs failed candidates through LLM-generated grammars for
    constrained symbolic synthesis.
    
    \item We demonstrate on seven RTL benchmarks that the full
    framework consistently improves both assertion quantity and mutation
    coverage over SMART and direct LLM generation. \looseness = -1
\end{itemize}


%% file: sections/background.tex
\section{Preliminaries}
\label{sec:preliminaries}

We consider RTL designs written in Verilog/SystemVerilog, with target
properties expressed as SystemVerilog Assertions (SVAs). This section
reviews the two ingredients underlying our methodology: SyGuS-based
assertion mining and mutation-coverage feedback.

\subsection{SyGuS-Based Assertion Mining}
Syntax-Guided Synthesis (SyGuS) synthesizes programs that satisfy a
specification under a context-free grammar~\cite{alur2013syntax}. Given
behavioral constraints $\varphi$
and a grammar $G = (\mathit{NT}, \mathit{T}, S, R)$, where $\mathit{NT}$
is the set of nonterminals, $\mathit{T}$ is the set of terminals, $S$
is the start symbol, and $R$ is the set of production rules, SyGuS searches for
formula $P$ derivable from $G$ such that $P$ satisfies $\varphi$.
By restricting candidates to those generated by $G$, SyGuS turns
unconstrained synthesis into structured search over a
grammar-defined space.

Prior work~\cite{ye2025unlocking} shows that this formulation is
effective for RTL assertion mining: traces provide the behavioral
constraints, while the grammar specifies the allowed operators, signal
combinations, and temporal structures of the synthesized SVAs. To keep
search tractable, existing methods partition RTL variables into smaller
groups, assign a predefined grammar to each group, and synthesize
assertions over each local signal set. Counterexamples from failed
assertions are then fed back through a counterexample-guided abstraction
refinement (CEGAR) loop to refine the trace-derived constraints. Our
work inherits this SyGuS-centered formulation, but strengthens the
behavioral basis of mining through formal exploration and extends
refinement beyond the initial SyGuS loop.

\subsection{Mutation Coverage and Uncovered Obligations}
In our framework, mutation coverage serves not only as an evaluation
metric but also as a refinement signal. We inject mutants that perturb
RTL operators, conditions, or assignments, and measure whether the
current assertion set can distinguish them from the original design
through formal checking. Mutants that remain undistinguished expose
behaviors insufficiently constrained by the current assertions. We
treat the resulting missing constraints as \emph{uncovered
obligations}: explicit verification targets that connect coverage to
refinement. Rather than treating incompleteness as an abstract problem,
we localize it into specific uncovered behaviors that guide downstream
generation and repair. In our methodology, uncovered obligations trigger
targeted neural proposal, while formal checking determines whether a
candidate can be accepted directly or must be repaired symbolically. \looseness = -1

%% file: sections/Method.tex
\section{Methodology}
\label{sec:method}

\subsection{Challenges to Address}

\subsubsection{C1: Limited Behavioral Reachability}
Despite improvements in synthesis and template design, most assertion
mining workflows still rely on random or constrained-random traces as
their initial behavioral basis. This makes mining fundamentally
reachability-limited: if important RTL behaviors are never exercised, no
downstream mining step can recover them. Hard-to-reach conditions are
especially problematic. For example, a branch condition such as
\texttt{a == 8'd222} over an 8-bit signal is triggered with probability
only 1/256 in each random test.

\begin{figure}[t]
    \centering
    \includegraphics[width=1\linewidth]{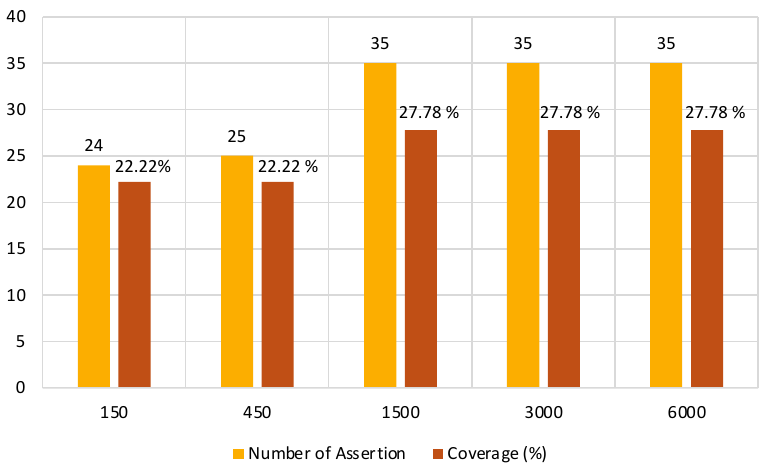}
    \caption{Coverage and assertion saturation with increasing random
    test cases on the b12 benchmark. Despite a 40$\times$ increase in
    simulation effort, branch coverage plateaus at about 27\% and the
    assertion count saturates at 35.}
    \label{fig:motivating_example}
    \vspace{-3mm}
\end{figure}

More importantly, simply increasing the number of random simulations
does not remove this bottleneck. As shown in
\cref{fig:motivating_example}, even when using the state-of-the-art
SMART method~\cite{ye2025unlocking} and scaling from 150 to 6000 test
cases on the b12 benchmark from the ITC'99
suite~\cite{davidson1999characteristics}, both branch coverage and the
number of mined assertions plateau at around 28\% and 35,
respectively. This saturation indicates that random exploration reaches
a ceiling on certain designs, leaving potentially important RTL
behaviors unobserved and limiting the completeness of the mined
assertion set.

\subsubsection{C2: Weak Refinement Feedback}
Even after an initial assertion set has been mined, existing workflows
provide limited guidance about how the assertion set should be expanded.
They mainly validate whether the current assertions hold, but seldom
identify which design behaviors remain uncovered or elevate these gaps
into explicit refinement targets for the next iteration. As a result,
the available feedback is largely about the correctness of existing
assertions rather than the completeness of the assertion set. This
stands in contrast to practical verification workflows, where engineers
often inspect coverage feedback to decide what assertions should be
added next. Existing automated methods lack a similarly directed
refinement signal.

\subsection{Overview of NeuroAssertion}
The above challenges suggest treating RTL assertion mining as a
coverage-driven refinement loop rather than a one-shot generation
process. To address C1, NeuroAssertion introduces S1, a formal
exploration stage that converts hard-to-reach control-flow conditions
into reachability objectives and uses the resulting traces to
strengthen SyGuS-based initial mining. To address C2, it introduces
S2, a feedback-driven neuro-symbolic refinement stage that uses
mutation coverage to identify uncovered obligations, propose candidate
assertions, and repair candidates that fail checking through
grammar-constrained symbolic synthesis.

\begin{figure*}[t]
    \centering
    \includegraphics[width=0.86\linewidth]{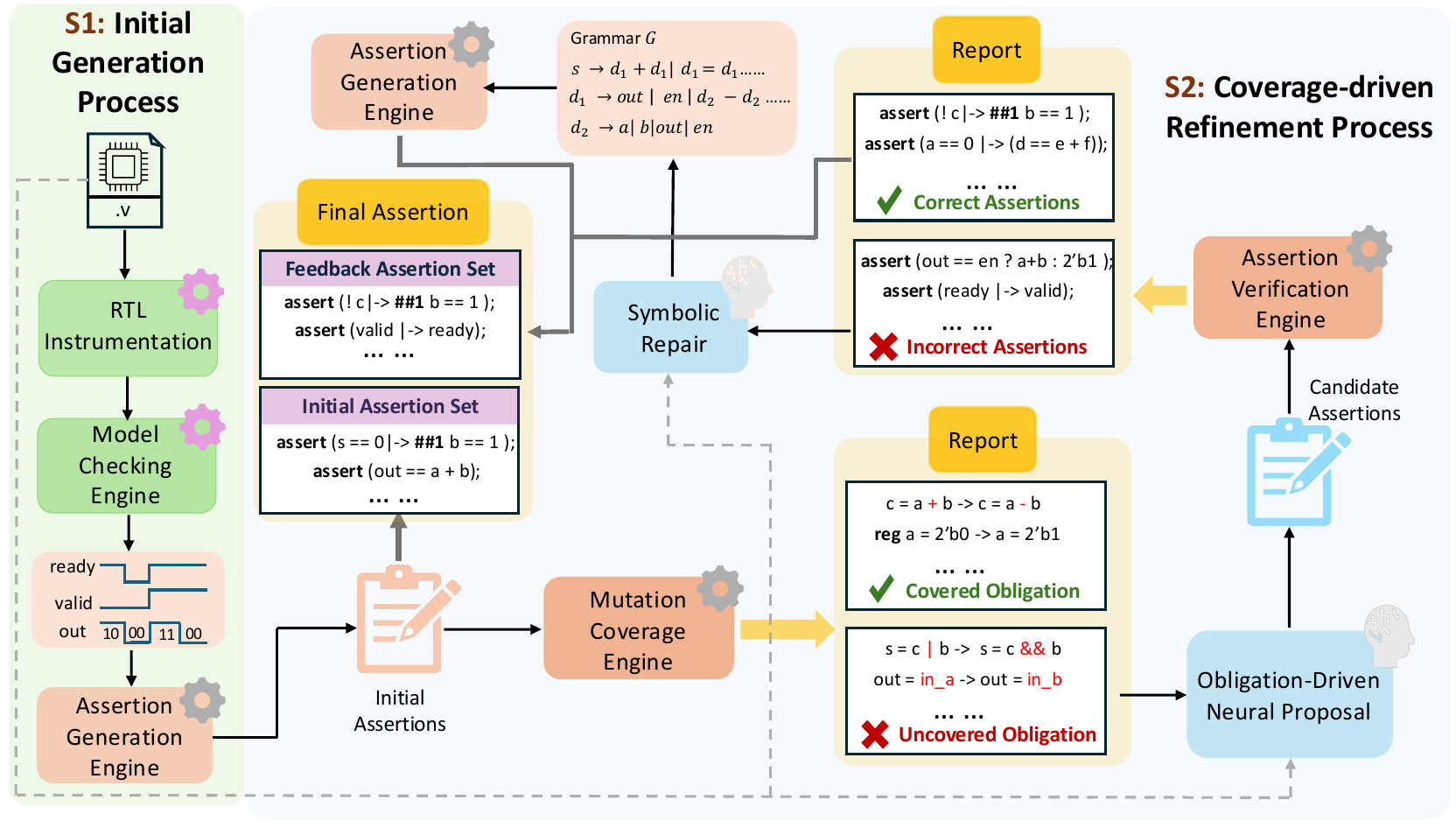}
    \caption{Overview of NeuroAssertion. S1 performs initial assertion
    generation, while S2 performs feedback-driven refinement. The final
    assertion set combines initial, directly verified, and repaired
    assertions.}
    \label{fig:overview}
    \vspace{-1mm}
\end{figure*}

\cref{fig:overview} summarizes the pipeline. RTL instrumentation and
model checking first generate traces that strengthen SyGuS-based mining
and produce an initial assertion set $\mathcal{A}_0$. Mutation analysis
then identifies uncovered obligations and drives a refinement cycle of
neural proposal, formal checking, and symbolic repair. Candidates that
pass checking are added directly to the assertion set, while candidates
that fail are repaired symbolically. In this way, mutation analysis
provides completeness feedback, and formal verification provides
correctness feedback for targeted refinement. \looseness = -1

\subsection{S1: Formal Exploration for Enhanced Assertion Mining}
To address C1, S1 strengthens assertion mining before refinement begins.
The key idea is to expand the behavioral basis of mining by actively
seeking executions that random simulation is unlikely to expose.
Prior work has shown that formal reachability analysis can
systematically drive execution toward target behaviors that are hard to
exercise through random simulation~\cite{zheng2025hot}. Following this
intuition, we first perform RTL instrumentation for targeted
reachability analysis, converting target control-flow conditions into
explicit formal reachability objectives. We then use model checking to
search for executions that satisfy these objectives. In this way, the
generated traces expose a broader range of execution patterns and
provide a stronger basis for downstream assertion synthesis.

\begin{figure}[t]
    \centering
    \includegraphics[width=0.48\linewidth]{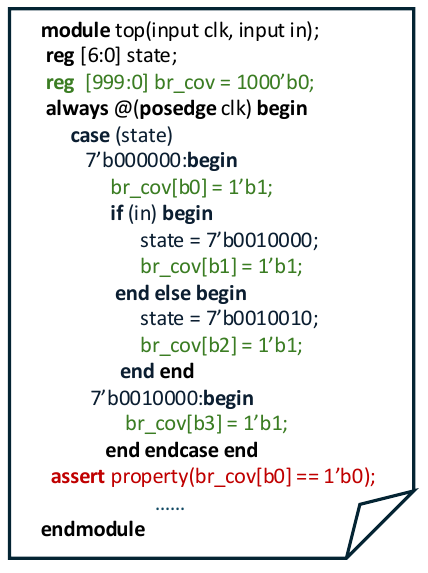}
    \caption{RTL instrumentation for targeted reachability analysis. Inserted code
    is color-coded: auxiliary registers and branch markers are
    highlighted in green, while the generated SVA is highlighted in
    red.}
    \label{fig:RTL}
    \vspace{-3mm}
\end{figure}

\cref{fig:RTL} shows this instrumentation process. For each branch
condition in the RTL, we introduce zero-initialized auxiliary coverage
registers \texttt{br\_cov} whose bits record whether the corresponding
branches have been exercised, and we insert update statements that set
the relevant bit when the target branch is taken. We then generate SVAs
over these coverage bits as formal objectives. Counterexample traces
returned by the model checker become test vectors that exercise the
target branches.

After obtaining these traces, we run SyGuS within an iterative CEGAR
loop to synthesize an initial assertion set $\mathcal{A}_0$ over the
observed behaviors. Counterexamples returned by formal verification are
used to refine the trace-derived constraints and guide SyGuS toward
improved assertions. S1 therefore addresses C1 not by changing the
symbolic miner itself, but by giving it a richer set of behaviors to
mine from.

\subsection{S2: Coverage-Driven Neuro-Symbolic Refinement}
To address C2, S2 uses mutation coverage to derive a set of uncovered
obligations $\mathcal{U}_0$ from the initial assertion set
$\mathcal{A}_0$ and uses them to drive refinement through two tightly
coupled stages: an
obligation-driven neural proposal stage and a grammar-constrained
symbolic repair stage.

\subsubsection{Obligation-Driven Neural Proposal}
Given RTL context and uncovered obligations $\mathcal{U}_0$, the first
LLM generates a candidate set $\tilde{\mathcal{A}}$. The role of this
stage is not to certify correctness, but to provide semantically
informed proposals for regions that remain uncovered after the initial
symbolic mining stage. In this sense, it serves as the neural front end
of the refinement loop: the LLM expands the candidate space for
hard-to-cover behaviors, while correctness is delegated to the formal
validation stages that follow.

To keep the generated candidates useful for downstream verification, we
use a structured prompting skill that guides the model through four steps:
\begin{itemize}
\item \textbf{Mutation localization and context extraction}: Identify
the mutation site, the changed expression, and the surrounding RTL
context.
\item \textbf{Circuit type detection}: Determine whether the design is
combinational or sequential, and identify the relevant clock and reset
signals.
\item \textbf{Behavioral intent inference}: Infer how the uncovered
obligation changes the intended behavior and express that behavior as
logical conditions.
\item \textbf{Signal extraction and assertion synthesis}: Extract the
relevant signals and emit synthesis-compatible SystemVerilog assertions
with metadata.
\end{itemize}


Each candidate is emitted as a structured JSON object containing the
assertion, its semantic description, the required signals, and
verification metadata. The prompt constrains outputs to
synthesis-compatible SystemVerilog assertions that can be checked by the
formal backend without manual rewriting. As a result, the proposal
stage integrates smoothly with the later verification and repair stages
instead of introducing an additional manual translation step.

\subsubsection{Grammar-Constrained Symbolic Repair}
Each candidate assertion proposed in the previous stage is checked
immediately by the formal backend. Valid assertions are inserted
directly into the current assertion set, while invalid ones are passed
to a repair stage rather than being discarded. The main reason is that
the variables involved in a failed candidate often still encode
meaningful behavioral relationships, even if the assertion itself is not
yet correct. Restarting generation from scratch would discard this
useful structure. \looseness = -1

For a failed candidate assertion $\tilde{a} \in \tilde{\mathcal{A}}_t$,
we extract its relevant signal set $S$ and use the second LLM to map the
RTL context $R$ and $S$ to a repair-oriented grammar:
\[
G = \mathrm{LLM}_2(R, S)
\]
This constrained grammar turns repair from an unconstrained symbolic
search under a general grammar into a local search centered on the
signals implicated by the failed candidate. The neural component
therefore does not emit a repaired assertion directly; instead, it
defines a semantically focused symbolic search space.

Once this grammar is constructed, the remaining repair process is
symbolic. We invoke SyGuS under the grammar $G$ and the constraint set
$C$ to synthesize a repaired assertion:
\[
a^{\star} = \mathrm{SyGuS}(G, C)
\]
If $a^{\star}$ still fails formal validation, we preserve the newly
observed counterexample and continue repair under strengthened
constraints. When the LLM-induced grammar is insufficient, we fall back
to a default grammar over the same signal set and continue
counterexample-guided symbolic synthesis.

\begin{algorithm}[t]
\caption{Coverage-driven refinement with neural proposal and symbolic repair}
\label{alg:repair_loop}
\scriptsize             
\KwIn{RTL context $R$, trace-derived constraint set $C_{\mathrm{trace}}$, current assertion set $\mathcal{A}$, uncovered obligations $\mathcal{U}$}
\KwOut{Refined assertion set $\mathcal{A}$}
\ForEach{$u \in \mathcal{U}$}{
  $\tilde{a} \leftarrow \mathrm{LLM}_1(R, u)$\; \label{alg:llm1}
  $c \leftarrow \mathrm{FormalCheck}(R, \tilde{a})$\; \label{alg:check1}
  \If{$c = \emptyset$}{
    $\mathcal{A} \leftarrow \mathcal{A} \cup \{\tilde{a}\}$\; \label{alg:add1}
    \Continue\;
  }
  $C \leftarrow C_{\mathrm{trace}} \cup \{c\}$\; \label{alg:initcex}
  $S \leftarrow \mathrm{ExtractSignals}(\tilde{a})$\; \label{alg:extract}
  $G \leftarrow \mathrm{LLM}_2(R, S)$\; \label{alg:llm2}
  $a^{\star}, c \leftarrow \mathrm{GrammarGuidedRepair}(R, G, C)$\; \label{alg:sygus1}
  \If{$c = \emptyset$}{
    $\mathcal{A} \leftarrow \mathcal{A} \cup \{a^{\star}\}$\; \label{alg:add2}
    \Continue\;
  }
  $C \leftarrow C \cup \{c\}$\; \label{alg:updatecex}
  $G \leftarrow \mathrm{DefaultGrammar}(S)$\; \label{alg:default}
  $a^{\star} \leftarrow \mathrm{DefaultRepair}(R, G, C)$\; \label{alg:sygus2}
  $\mathcal{A} \leftarrow \mathcal{A} \cup \{a^{\star}\}$\; \label{alg:add3}
}
\end{algorithm}

Algorithm~\ref{alg:repair_loop} summarizes the overall refinement
procedure. For each uncovered obligation, the first LLM proposes a
candidate assertion (line~\ref{alg:llm1}), and the formal backend
checks it immediately (line~\ref{alg:check1}). Candidates that already
satisfy formal checking are inserted directly into the assertion set
(line~\ref{alg:add1}).

If the initial candidate fails, the returned counterexample is added to
the trace-derived constraints from S1 to initialize the repair
constraint set (line~\ref{alg:initcex}), and the relevant signals are
extracted from the failed candidate (line~\ref{alg:extract}). The
second LLM then generates a repair grammar (line~\ref{alg:llm2}). The
grammar-guided repair solver runs counterexample-guided symbolic
synthesis under that grammar and returns either a valid repaired
assertion or a new counterexample (line~\ref{alg:sygus1}). If repair
succeeds, the repaired assertion is inserted into the assertion set
(line~\ref{alg:add2}). Otherwise, the new counterexample is added to the
constraint set (line~\ref{alg:updatecex}), the search switches to a
default grammar over the same signal set (line~\ref{alg:default}), and
a fallback symbolic repair solver is invoked (line~\ref{alg:sygus2}).
The resulting valid assertion is then added to the assertion set
(line~\ref{alg:add3}). In this way, the neural stages define the repair
space and the symbolic stages carry out counterexample-guided synthesis
and final certification.

%% file: sections/Experiment.tex
\section{Experiment}
\label{sec:experiment}
\subsection{Experimental Setup}
The experiments are conducted on a machine with a 2.9 GHz Intel
Xeon Platinum 8375C CPU and 256 GB RAM. For the initial
assertion-mining stage, we follow the variable-grouping heuristic
of~\cite{ye2025unlocking}, and we use cvc5~\cite{barbosa2022cvc5} as
the SyGuS solver for assertion
synthesis. For all LLM-related tasks in our framework, including
assertion proposal and grammar generation, we use GPT-5 as the
underlying large language model. For formal checking, we use
SymbiYosys~\cite{symbiyosys} to preprocess the design and translate it
into the BTOR2 format, and then use Pono~\cite{mann2021pono} to verify
whether the generated assertions hold on the design.

As the main baseline, we compare against SMART, the oracle-guided
SyGuS-based method proposed by Ye et al.~\cite{ye2025unlocking}. We
choose SMART because it is the most closely related prior work to our
approach, sharing the same SyGuS-centered assertion generation
framework. In contrast, our method further introduces
formal exploration and an LLM-guided refinement stage after the initial mining
stage. Earlier methods such as GoldMine~\cite{vasudevan2010goldmine}
and HARM~\cite{germiniani2022harm} represent important prior
approaches, but SMART already showed stronger results than these
baselines, making it the most appropriate primary comparison target in
our experiments.

\subsection{Benchmarks}
We evaluate our method on seven RTL benchmarks spanning a range of
design sizes and complexities. Arb2 is a compact arbitration benchmark
adopted from prior assertion-mining evaluation~\cite{ye2025unlocking},
and B12 is taken from the ITC'99 benchmark suite~\cite{davidson1999characteristics}.
I2C corresponds to the OpenCores I2C controller~\cite{opencoresi2c}.
To include processor-related RTL modules, we also evaluate three
components from the open-source Ibex RISC-V core~\cite{ibex}:
Ibex\_controller, Ibex\_decoder, and Multdiv. Finally, Pico is derived
from the PicoRV32 RISC-V core~\cite{picorv32}. Table~\ref{tab:benchmarks}
summarizes the statistics of these benchmarks. Overall, the benchmark
set covers designs ranging from 59 to 2530 RTL lines and from 13 to
759 variables, providing a heterogeneous testbed for evaluating
assertion generation across substantially different design sizes.

\begin{table}[t]
    \centering
    \caption{The statistics of benchmark designs in the experiment.}
    \renewcommand{\arraystretch}{1.0} 
\scalebox{0.99}{
\setlength{\tabcolsep}{1mm}
{
    \label{tab:benchmarks}
    \begin{tabular}{|c|c|c|}
        \hline
        Design & \# RTL Lines & \# Variables \\
        \hline
        \hline
        Arb2             &   59 &  13 \\
        \hline
        B12              &  738 &  26 \\
        \hline
        I2C              & 1114 & 155 \\
        \hline
        Ibex\_controller & 1101 & 233 \\
        \hline
        Ibex\_decoder    & 1156 & 141 \\
        \hline
        Multdiv          &  353 &  59 \\
        \hline
        Pico             & 2530 & 759 \\
        \hline
    \end{tabular}
    }
    }
\end{table}

\subsection{Experimental Result}
\begin{figure}[t]
    \centering
    \includegraphics[width=0.93\linewidth]{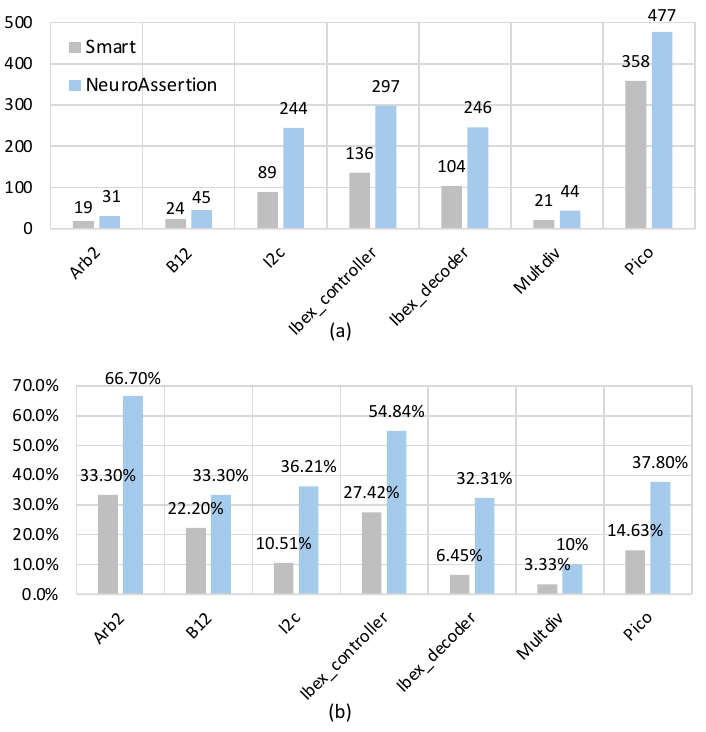}
    \caption{Overall experimental results compared with SMART. (a) shows the number
    of generated assertions, and (b) shows mutation
    coverage rate.}
    \label{fig:result_with_smart}
\end{figure}

\Cref{fig:result_with_smart} presents the overall experimental results
of our method and SMART~\cite{ye2025unlocking} on the benchmark suite.
Overall, our method achieves stronger performance across the
benchmarks, indicating that the proposed coverage-driven refinement
pipeline improves the quality of the generated assertion set beyond a
purely SyGuS-based workflow. In terms of assertion quantity, our method
generally produces around twice as many assertions as SMART. The gain
is especially pronounced on larger designs such as I2C (89 vs. 244),
Ibex\_controller (136 vs. 297), Ibex\_decoder (104 vs. 246), and Pico
(358 vs. 477). In terms of mutation coverage, our method also delivers
around a twofold improvement overall, and on several benchmarks the
gain is even larger, including I2C (10.51\% to 36.21\%),
Ibex\_decoder (6.45\% to 32.31\%), and Pico (14.63\% to 37.80\%).
These results show that our method not only generates more assertions,
but also produces assertion sets that detect substantially more
mutations.

\subsection{Ablation Study}
\begin{figure}[t]
    \centering
    \includegraphics[width=1.02\linewidth]{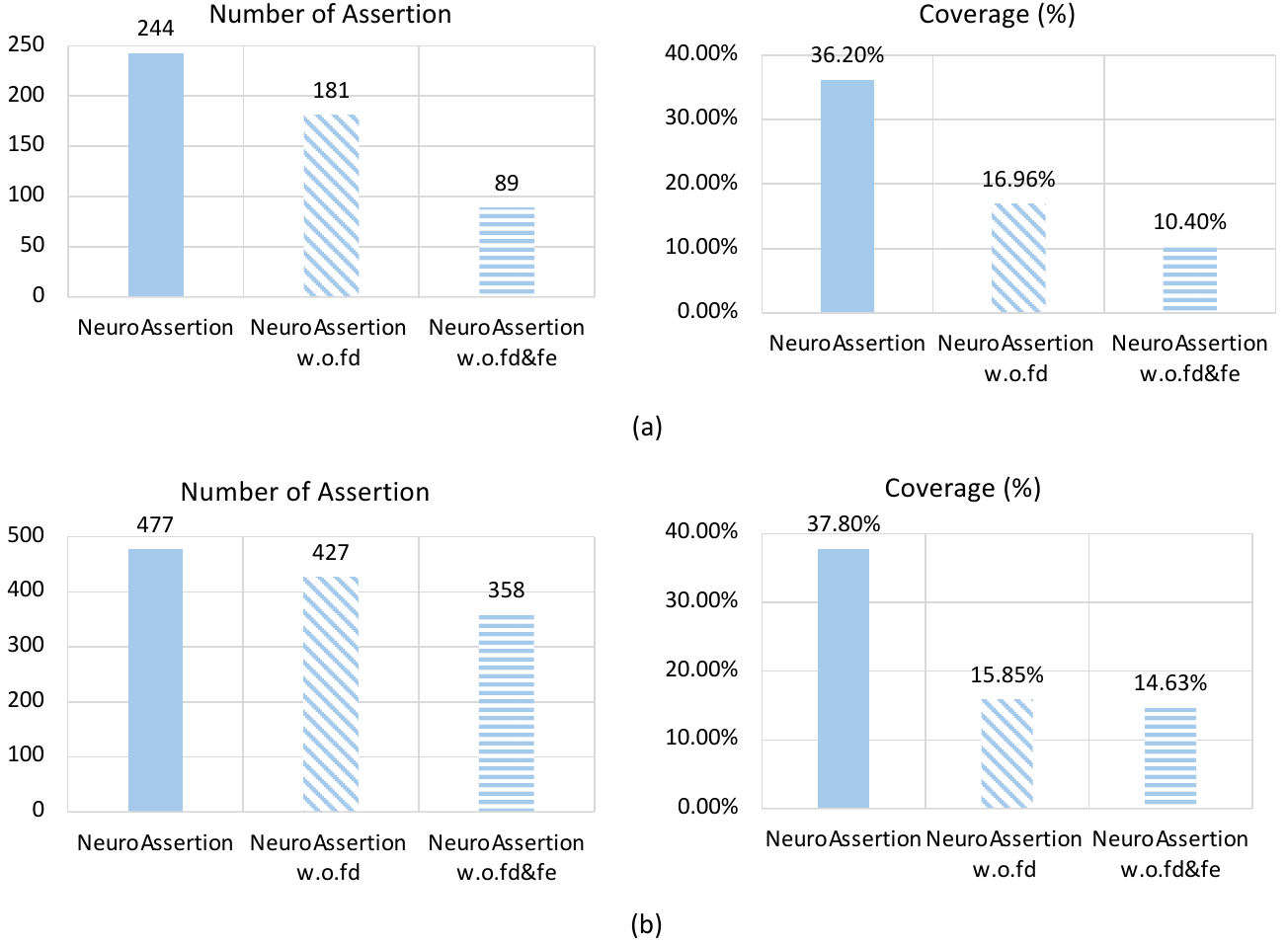}
    \vspace{-2mm}
    \caption{Ablation study of NeuroAssertion on representative
    benchmarks. (a) I2C, and (b) Pico.}
    \label{fig:ablation}
\end{figure}

\Cref{fig:ablation} further studies the contribution of the formal
exploration and feedback components. We compare the full
NeuroAssertion framework with two reduced variants:
NeuroAssertion w.o.fd and NeuroAssertion w.o.fd\&fe, which remove the
feedback component (fd) and both the formal-exploration and
feedback components (fd\&fe), respectively. We report results on two
representative benchmarks, I2C and Pico.

The ablation results show that both components are important to the
final performance. On I2C, the full method generates 244 assertions and
achieves 36.20\% mutation coverage, while NeuroAssertion w.o.fd drops
to 181 assertions and 16.96\% coverage, and NeuroAssertion
w.o.fd\&fe further drops to 89 assertions and 10.40\% coverage. A
similar trend appears on Pico, where the full method generates 477
assertions with 37.80\% coverage, compared with 427 assertions and
15.85\% coverage for NeuroAssertion w.o.fd, and 358 assertions and
14.63\% coverage for NeuroAssertion w.o.fd\&fe.

These results indicate that formal exploration is essential for
exposing richer design behaviors and increasing the number of useful
candidate assertions, while the refinement process is
critical for turning these candidates into assertions that improve
mutation coverage. Together, these components make the full
NeuroAssertion framework substantially more effective than its reduced
variants.

\subsection{Comparison with Direct LLM Generation}
\begin{figure}[t]
    \centering
    \includegraphics[width=0.93\linewidth]{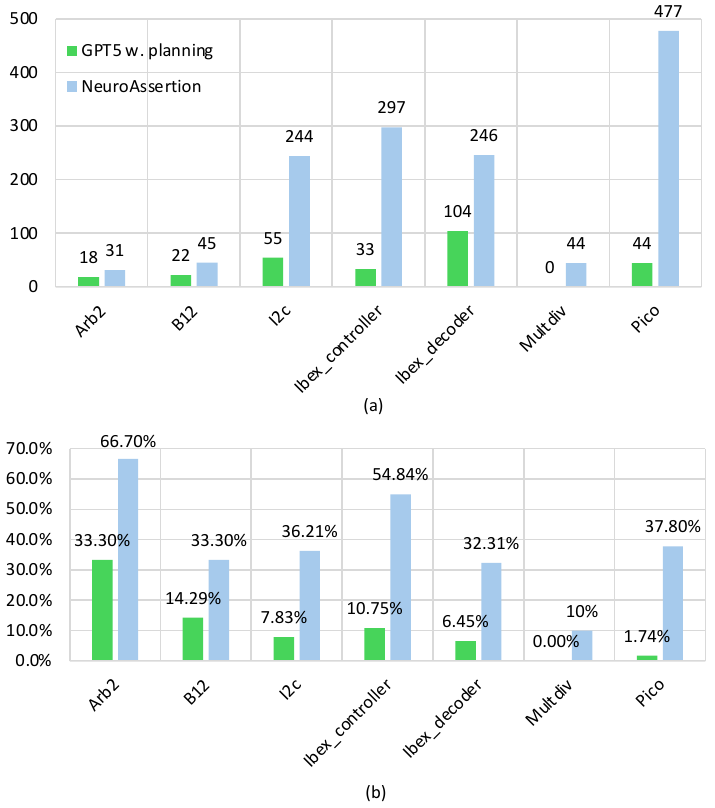}
    \caption{Comparison with a direct LLM-based baseline. Subfigure
    (a) shows the number of generated assertions, and subfigure (b)
    shows mutation coverage.}
    \label{fig:result_with_llm}
\end{figure}

Existing assertion-mining methods,\footnote{We exclude other
RTL-oriented methods because they target different settings:
Orenes-Vera et al.~\cite{orenesvera2023llmrtl} involves iterative
manual prompt engineering, while
Yan and Zhang~\cite{yan2025wordlevel} focuses on accelerating model
checking rather than automatic assertion generation.} including our
main comparison with
SMART, do not directly evaluate whether an LLM alone can generate
assertions from RTL code. To study this setting, we construct a direct
LLM baseline, denoted GPT-5 w. planning. This baseline uses GPT-5 with
task-specific prompts that first analyze the RTL structure and then
generate assertions for connectivity, register behavior, function,
constraints, and safety. In this way, the baseline still benefits from
prompt-level decomposition and planning, but it does not use our formal
exploration stage or feedback-driven neuro-symbolic refinement loop.

\Cref{fig:result_with_llm} shows that NeuroAssertion substantially
outperforms this direct LLM baseline on both assertion quantity and
mutation coverage. In terms of assertion quantity, NeuroAssertion
generally produces several times more assertions on complex benchmarks,
including I2C (244 vs. 55), Ibex\_controller (297 vs. 33), Pico
(477 vs. 44), and Multdiv (44 vs. 0). In terms of mutation coverage,
the gap is even larger: NeuroAssertion improves coverage from 7.83\% to
36.21\% on I2C, from 10.75\% to 54.84\% on Ibex\_controller, from
1.74\% to 37.80\% on Pico, and from 0.00\% to 10.00\% on Multdiv.
Even on benchmarks where GPT-5 w. planning can generate a nontrivial
number of assertions, such as Arb2 and Ibex\_decoder, NeuroAssertion
still achieves much higher coverage.

These results show that prompt-level planning alone is insufficient for
high-quality RTL assertion generation. Although GPT-5 w. planning can
produce some plausible assertions directly from RTL structure, without
formal exploration it misses many hard-to-reach behaviors, and without
our refinement loop it cannot reliably repair weak or incorrect
candidates. NeuroAssertion therefore provides a much more effective way
to turn LLM reasoning into assertions that are both numerous and
verification-useful.

%% file: sections/related-works.tex
\section{Related Works}
\label{sec:related-works}

Recent machine-learning-based methods for hardware verification can be
broadly grouped into methods driven primarily by
specification-side artifacts and RTL-oriented methods. Methods in the
former category are driven by natural-language requirements, waveform
descriptions, or coverage targets, including
NL2SVA~\cite{sun2023nl2sva}, ChatSVA~\cite{fu2026chatsva},
ChIRAAG~\cite{mali2024chiraag}, AssertLLM~\cite{yan2025assertllm},
CoverAssert~\cite{wang2026coverassert}, and
LASP~\cite{ayalasomayajula2024lasp}. Representative RTL-oriented
methods include AutoSVA~\cite{orenesvera2021autosva}, the RTL-focused
LLM-assisted flow of Orenes-Vera et al.~\cite{orenesvera2023llmrtl},
and the trace-based word-level assertion generation method  DreamMiner~\cite{yan2025wordlevel}. However, the former still involves
iterative manual prompt engineering, while the latter focuses on
accelerating model checking rather than automatic assertion generation.
Our method belongs to the
RTL-oriented category, but its focus is to automatically refine
assertions through a feedback-driven loop that combines LLM-based
generation with traditional assertion-mining and formal-checking tools
under RTL-grounded mutation coverage. \looseness=-1

%% file: sections/conclusion.tex
\section{Conclusion}
\label{sec:conclusion}

This paper presented NeuroAssertion, a coverage-driven framework for
RTL assertion generation that combines formal exploration,
syntax-guided synthesis, and a feedback-driven neuro-symbolic
refinement process. NeuroAssertion first expands the behavioral basis of
assertion mining through RTL instrumentation and model checking, and
then uses mutation-coverage feedback to identify uncovered obligations
that guide targeted neural proposal and symbolic repair. In this way,
the framework turns mutation analysis from a post hoc evaluation metric
into an explicit refinement signal for assertion generation.
Experiments further show that NeuroAssertion consistently outperforms
SMART and a direct LLM baseline, producing around 2$\times$ more
assertions and achieving about 2$\times$ higher mutation coverage
overall.

%% file: sample-base.bib
@book{alur2013syntax,
  title={Syntax-guided synthesis},
  author={Alur, Rajeev and Bodik, Rastislav and Juniwal, Garvit and Martin, Milo MK and Raghothaman, Mukund and Seshia, Sanjit A and Singh, Rishabh and Solar-Lezama, Armando and Torlak, Emina and Udupa, Abhishek},
  year={2013},
  publisher={IEEE}
}

@inproceedings{mann2021pono,
  title={Pono: a flexible and extensible SMT-based model checker},
  author={Mann, Makai and Irfan, Ahmed and Lonsing, Florian and Yang, Yahan and Zhang, Hongce and Brown, Kristopher and Gupta, Aarti and Barrett, Clark},
  booktitle={Computer Aided Verification: 33rd International Conference, CAV 2021, Virtual Event, July 20--23, 2021, Proceedings, Part II 33},
  pages={461--474},
  year={2021},
  organization={Springer}
}

@inproceedings{barbosa2022cvc5,
  title={cvc5: A versatile and industrial-strength SMT solver},
  author={Barbosa, Haniel and Barrett, Clark and Brain, Martin and Kremer, Gereon and Lachnitt, Hanna and Mann, Makai and Mohamed, Abdalrhman and Mohamed, Mudathir and Niemetz, Aina and Ozdemir, Alex and N{\"o}tzli, Andres and Preiner, Mathias and Reynolds, Andrew and Sheng, Ying and Tinelli, Cesare and Zohar, Yoni},
  booktitle={Tools and Algorithms for the Construction and Analysis of Systems},
  pages={415--442},
  year={2022},
  organization={Springer}
}

@misc{symbiyosys,
  author={{YosysHQ}},
  title={SymbiYosys (SBY) Documentation},
  year={2026},
  howpublished={\url{https://symbiyosys.readthedocs.io/en/latest/}},
  note={Accessed: 2026-04-12}
}

@misc{ibex,
  author={{lowRISC}},
  title={Ibex RISC-V Core},
  year={2026},
  howpublished={\url{https://github.com/lowRISC/ibex}},
  note={Accessed: 2026-04-14}
}

@misc{picorv32,
  author={{YosysHQ}},
  title={PicoRV32: A Size-Optimized RISC-V CPU},
  year={2026},
  howpublished={\url{https://github.com/YosysHQ/picorv32}},
  note={Accessed: 2026-04-14}
}

@misc{opencoresi2c,
  author={Herveille, Richard},
  title={OpenCores I2C Controller Core},
  year={2026},
  howpublished={\url{https://opencores.org/projects/i2c}},
  note={Accessed: 2026-04-14}
}

@inproceedings{vasudevan2010goldmine,
  title={Goldmine: Automatic assertion generation using data mining and static analysis},
  author={Vasudevan, Shobha and Sheridan, David and Patel, Sanjay and Tcheng, David and Tuohy, Bill and Johnson, Daniel},
  booktitle={2010 Design, Automation \& Test in Europe Conference \& Exhibition (DATE 2010)},
  pages={626--629},
  year={2010},
  organization={IEEE}
}

@inproceedings{liu2011automatic,
  title={Automatic generation of assertions from system level design using data mining},
  author={Liu, Lingyi and Sheridan, David and Athavale, Viraj and Vasudevan, Shobha},
  booktitle={Ninth ACM/IEEE International Conference on Formal Methods and Models for Codesign (MEMPCODE2011)},
  pages={191--200},
  year={2011},
  organization={IEEE}
}

@inproceedings{liu2012word,
  title={Word level feature discovery to enhance quality of assertion mining},
  author={Liu, Lingyi and Lin, Chen-Hsuan and Vasudevan, Shobha},
  booktitle={Proceedings of the International Conference on Computer-Aided Design},
  pages={210--217},
  year={2012}
}

@inproceedings{sheridan2014coverage,
  title={A coverage guided mining approach for automatic generation of succinct assertions},
  author={Sheridan, David and Liu, Lingyi and Kim, Hyungsul and Vasudevan, Shobha},
  booktitle={2014 27th International Conference on VLSI Design and 2014 13th International Conference on Embedded Systems},
  pages={68--73},
  year={2014},
  organization={IEEE}
}

@inproceedings{danese2017team,
  title={A-team: Automatic template-based assertion miner},
  author={Danese, Alessandro and Riva, Nicol{\`o} Dalla and Pravadelli, Graziano},
  booktitle={Proceedings of the 54th Annual Design Automation Conference 2017},
  pages={1--6},
  year={2017}
}

@article{germiniani2022harm,
  title={Harm: a hint-based assertion miner},
  author={Germiniani, Samuele and Pravadelli, Graziano},
  journal={IEEE Transactions on Computer-Aided Design of Integrated Circuits and Systems},
  volume={41},
  number={11},
  pages={4277--4288},
  year={2022},
  publisher={IEEE}
}

@inproceedings{ye2025unlocking,
  title={Unlocking hardware verification with oracle guided synthesis},
  author={Ye, Leiqi and Li, Yixuan and Frankel, Guy and Cheng, Jianyi and Polgreen, Elizabeth},
  booktitle={The 25th Conference on Formal Methods in Computer-Aided Design},
  pages={235--245},
  year={2025},
  organization={TU Wien Academic Press}
}

@inproceedings{davidson1999characteristics,
  title={Characteristics of the ITC’99 benchmark circuits},
  author={Davidson, Scott},
  booktitle={IEEE International Test Synthesis Workshop (ITSW)},
  pages={87},
  year={1999}
}

@inproceedings{zheng2025hot,
  title={Hot-FV: A Semi-Formal Test Generation Framework for RTL Functional Coverage Using Warm Starting States},
  author={Zheng, Ziyue and Yan, Zhiyuan and Meng, Xiangchen and Hu, Guangyu and Zhang, Hongce and Lyu, Yangdi},
  booktitle={2025 IEEE 43rd International Conference on Computer Design (ICCD)},
  pages={298--305},
  year={2025},
  organization={IEEE}
}

@inproceedings{orenesvera2021autosva,
  title={AutoSVA: Democratizing formal verification of RTL module interactions},
  author={Orenes-Vera, Marcelo and Manocha, Aninda and Wentzlaff, David and Martonosi, Margaret},
  booktitle={2021 58th ACM/IEEE Design Automation Conference (DAC)},
  pages={535--540},
  year={2021},
  organization={IEEE}
}

@article{orenesvera2023llmrtl,
  title={Using Large Language Models to Facilitate Formal Verification of RTL},
  author={Orenes-Vera, Marcelo and Martonosi, Margaret and Wentzlaff, David},
  journal={arXiv preprint arXiv:2309.09437},
  year={2023}
}

@article{mali2024chiraag,
  title={ChIRAAG: ChatGPT Informed Rapid and Automated Assertion Generation},
  author={Mali, Bhabesh and Maddala, Karthik and Reddy, Sweeya and Gupta, Vatsal and Karfa, Chandan and Karri, Ramesh},
  journal={arXiv preprint arXiv:2402.00093},
  year={2024}
}

@inproceedings{sun2023nl2sva,
  title={Towards Improving Verification Productivity with Circuit-Aware Translation of Natural Language to SystemVerilog Assertions},
  author={Sun, Chuyue and Hahn, Christopher and Trippel, Caroline},
  booktitle={Design Automation Conference Young Fellows Workshop (DAV)},
  year={2023},
  note={OpenReview}
}

@inproceedings{ayalasomayajula2024lasp,
  title={LASP: LLM Assisted Security Property Generation for SoC Verification},
  author={Ayalasomayajula, Avinash and Guo, Rui and Zhou, Jingbo and Saha, Sujan Kumar and Farahmandi, Farimah},
  booktitle={2024 ACM/IEEE 6th Workshop on Machine Learning for CAD (MLCAD)},
  pages={29:1--29:7},
  year={2024},
  doi={10.1145/3670474.3685967}
}

@inproceedings{yan2025assertllm,
  title={AssertLLM: Generating Hardware Verification Assertions from Design Specifications via Multi-LLMs},
  author={Yan, Zhiyuan and Fang, Wenji and Li, Mengming and Li, Min and Liu, Shang and Xie, Zhiyao and Zhang, Hongce},
  booktitle={30th Asia and South Pacific Design Automation Conference},
  pages={614--621},
  year={2025}
}

@inproceedings{yan2025wordlevel,
  title={Word-Level Augmentation of Formal Proof by Learning from Simulation Traces},
  author={Yan, Zhiyuan and Zhang, Hongce},
  booktitle={2024 IEEE/ACM International Conference on Computer-Aided Design (ICCAD)},
  pages={1--8},
  year={2024},
  doi={10.1145/3676536.3676686}
}

@article{wang2026coverassert,
  title={CoverAssert: Iterative LLM-Based Assertion Generation Using Syntax-Semantic Representations for Functional Coverage-Guided Verification},
  author={Wang, Yonghao and Zhou, Jiaxin and Yin, Yang and Lyu, Hongqin and Chao, Zhiteng and Ding, Wenchao and Ye, Jing and Wang, Tiancheng and Li, Huawei},
  journal={arXiv preprint arXiv:2602.15388},
  year={2026}
}

@article{fu2026chatsva,
  title={ChatSVA: Bridging SVA Generation for Hardware Verification via Task-Specific LLMs},
  author={Fu, Lik Tung and Zhou, Jie and Ren, Shaokai and Zhang, Mengli and Xiong, Jia and Jiang, Hugo and Guan, Nan and Wang, Xi and Yang, Jun},
  journal={arXiv preprint arXiv:2604.02811},
  year={2026}
}
